\documentclass[preprint2]{aastex}

\usepackage{amsmath}
\usepackage{import}
\usepackage{graphicx}
\usepackage{threeparttable} % footnotes in tables
\usepackage{booktabs} % tables
\usepackage{array} %tables

\usepackage{txfonts}
\usepackage{textcmds}
\usepackage{natbib}
\bibpunct{(}{)}{;}{a}{}{,}

\usepackage[bookmarks]{hyperref} % uses pdf bookmarks (sections, subsections, figure, cite, equation)

\usepackage{color} % to load*.eps_tex since it uses inkscape colors

\newcolumntype{L}[1]{>{\raggedright\arraybackslash}p{#1}} % flush left with width specification
\newcolumntype{C}[1]{>{\centering\arraybackslash}m{#1}} % centered with width specification
\newcolumntype{R}[1]{>{\raggedleft\arraybackslash}p{#1}} % flush right with width specification
\shorttitle{Photophoretic Strength on Chondrules 1: Modeling}
\shortauthors{Loesche et al.}

\begin{document}

   \title{PHOTOPHORETIC STRENGTH ON CHONDRULES.\\1. MODELING}

   \author{
          Christoph Loesche %\altaffilmark{1}
          and Gerhard Wurm %\altaffilmark{1}
          and Jens Teiser %\altaffilmark{1}
          }
   \affil{Faculty of physics, University of Duisburg-Essen, Lotharstr. 1, D-47057 Duisburg, Germany}
   \email{christoph.loesche@uni-due.de}
   
   \author{Jon M. Friedrich } %\altaffilmark{2,3}}
   \affil{Department of Chemistry, Fordham University, Bronx, NY 10458, USA
              and Department of Earth and Planetary Sciences, American Museum of Natural History, New York, NY 10024, USA}
   
   \author{Addi Bischoff} %\altaffilmark{4} }
   \affil{Institut f\"ur Planetologie, Westf\"alische Wilhelms-Universit\"at M\"unster, Wilhelm-Klemm-Str. 10, D-48149 M\"unster, Germany}

%   \date{ Received \today; accepted }
   
   \begin{abstract}
         Photophoresis is a physical process that transports particles in optical thin parts of protoplanetary disks, especially at the inner edge and at the optically surface. To model the transport and resulting effects in detail, it is necessary to quantify the strength of photophoresis for different particle classes as a fundamental input. Here, we explore photophoresis for a set of chondrules. The composition and surface morphology of these chondrules was measured by X-ray tomography. Based on the three-dimensional models, heat transfer through illuminated chondrules was calculated. The resulting surface temperature map was then used to calculate the photophoretic strength. 
         We found that irregularities in particle shape and variations in composition induce variations in the photophoretic force. These depend on the orientation of a particle with respect to the light source. The variations of the absolute value of the photophoretic force on average over all chondrules is $4.17$\%. The deviation between the direction of the photophoretic force and illumination is $3.0^\circ \pm 1.5^\circ$.
         The average photophoretic force can be well approximated and calculated analytically assuming a homogeneous sphere with a volume equivalent mean radius and an effective thermal conductivity.
         We found an analytic expression for the effective thermal conductivity. The expression depends on the two main phases of a chondrule, and decreases with the amount of fine-grained devitrified, plagioclase-normative mesostasis up to factor of three. For the chondrule sample studied (Bjurb\"ole chondrite), we found a dependence of the photophoretic force on chondrule size.
      \end{abstract}

    % source: http://www.aanda.org/index2.php?option=com_content&task=view&id=170&Itemid=256

%---eigentlicher Text der Veröffentlichung--------------------------------------------------------------------------------
\section{INTRODUCTION}

It is widely accepted that disks consisting of solids and gas give birth to planetary systems. However, many aspects of the formation process are still poorly understood.
Some constraints come from the analysis of matter from the solar system in the form of primitive chondritic meteorites.

In chondrites, components with different histories can be found. Calcium--aluminum-rich inclusions (CAIs), condensed directly from the gas phase, are among the oldest materials radiometrically dated due to their high condensation temperature \citep{Amelin2002, Wadhwa2007}.
Chondrules, major components of chondrites , likely formed by the melting of dust agglomerates. Although it is still not clear which processes lead to chondrule formation, it is well known that chondrules formed within a few million years after the first CAIs  \citep{Scott2007}.
Chondrules and CAIs are often found together in the same chondrite, embedded in a fine-grained matrix consisting of silicate dust.

The non-homogeneous composition of many meteorites is evidence for radial mixing of solid particles in the early  solar system.
Particles forming at different times and in different places within the solar nebula can be found in the same meteorite.
However, the complementarity of matrix and chondrules, the fact that volatile elements depleted in chondrules are enriched in the matrix restricts the relative transport of chondrules and dust \citep{Klerner1999, Hezel2010}.
Evidence for more local transport can be inferred  from the existence of meteorites with chondrules of different sizes, which implies the operation of a size sorting mechanism \citep{Kuebler1999, Scott1996, Hughes1978, Cuzzi1996, Liffman2005}.
One example could be the chondrules of the carbonaceous, geochemically related CR--CH--CB--chondrite clan, whose chondrule sizes vary considerably from $<$100 $\mu$m (CH-chondrites) up to about cm-sized in CB-chondrites \citep{Weisberg1995, Weisberg2010, Bischoff1993a, Bischoff1993b, Krot2005}.

Cometary studies give further evidence for radial transport in the early solar system.
Comets largely consist of ice and, therefore, must have formed in the outer parts of the solar system. In several comets, refractory minerals have been found by remote sensing \citep{Sitko2004, Wooden2004}.
Samples of comet Wild 2, which were collected by the \textit{Stardust} spacecraft, also contain high-temperature materials, which likely formed close to the Sun \citep{Zolensky2006, Brownlee2006}.

Various scenarios have been proposed to explain this radial redistribution and transport.
The transport of solid particles is strongly coupled to the gas dynamics in protoplanetary disks. Turbulence is a driving motor for gas flow \citep{Bockelee-Morvan2002, Cuzzi2003}.
As small particles couple well to gas movements, this leads to a random radial redistribution of solid particles.
Some models, taking into account the vertical disk structure, describe a radial outflow of particles due to pressure gradients within protoplanetary disks \citep{Ciesla2007, Keller2004}.
Other models use additional processes to explain the radial outbound movement of solid particles.
For example the X-wind model is based on ionized gas, which couples to the magnetic field of the central star and the inner disk. Particles in the direct environment of the star can be driven up and outward \citep{Shu1996, Shu1997}.
All transport models strongly depend on the assumed underlying disk model parameters.

The background behind this work is transport based on photophoresis, which has been introduced as a transport mechanism by \citet{Krauss2005} and \citet{Wurm2006}.

Photophoresis has been known since \citet{Fresnel1825} and is an interaction between illuminated solid particles and gas.
It requires only a gaseous environment and irradiation.
Due to illumination, a particle surface is heated on the side facing the light source and a temperature gradient develops.
Near the surface, the momentum of surrounding gas molecules changes according to the surface temperature of the particle.
Hence, the gas molecules near the warmer part acquire a larger thermal velocity than gas molecules on the cooler side of the particle.
This leads to a momentum transfer from the gas to the particle, which accelerates the particle along the direction of the temperature gradient in the direction from warm to cold.

In certain regions in protoplanetary disks, it is obligatory to consider photophoresis.
One  location is the surface of the disk. Both calculations and laboratory experiments demonstrate that particles can be photophoretically transported at higher levels of disks near its optical surface.
At this location, photophoresis can be induced by stellar radiation and thermal radiation from the disk \citep{Wurm2009,Eymeren2012}.

Another location where photophoretic forces are undoubtly at work is at the inner disk edge.
Early in the disk's history, this may be close to the sublimation radius but a number of transitional disks with large inner holes have also been observed \citep{Dalessio2005,Sicilia-Aguilar2008} where photophoretic particle transport should be considered \citep{Haack2007, Moudens2011}.
Further particle recycling by more complex photophoretic processes which are capable of disassembling larger dusty bodies can also take place at a disk's edge \citep{Wurm2007, Beule2013, Kelling2011}.
In several studies, it was shown that photophoresis can lead to enhanced particle concentrations and enhance growth rates at the inner edge of the dust disk \citep{Krauss2005, Wurm2006b}. Photophoresis  can also drive the inner edge of the dust disk further out \citep{Krauss2007, Moudens2011}. 
As photophoresis strongly depends on particle properties (e.g., thermal and optical properties), it is a general transport mechanism but can also lead to particle sorting by separation in the vicinity of the inner edge \citep{Wurm2006, Loesche2012, Wurm2013}.

Dust aggregates,  solid  particles like chondrules or CAIs, and dust mantled particles are three materials with differing properties \citep{Rohatschek1985,Steinbach2004,Borstel2012,Wurm2010}.
It was shown by \citet{Loesche2012} that sorting within a class is also possible, e.g., they studied the properties of dust mantled chondrules in detail.

This paper focuses on the photophoretic properties of bare chondrules without dust mantles.
Some previous experimental data considering photophoresis on chondrules exists but they show a large scatter, even for the same chondrule  \citep{Wurm2010,Hesse2011}.
Therefore, it is important to shed some light on the detailed influence of particle properties and temporal evolution.
Complex particle composition and particle shape are particularly major deviations from the usual assumption of material homogeneity and perfect sphericity.
Within this work, we study the influence of particle composition and shape on the photophoretic acceleration.
The three-dimensional composition of the chondrules is determined by X-ray tomography, and the heat transfer within the particles is modeled using Comsol Multiphysics.
The results are compared to calculations for chondrule-shaped homogeneous particles and spherical particles with typical chondrule-like compositions.
In a forthcoming paper (Paper 2) we will extend this work to a new set of microgravity experiments.
\section{PHOTOPHORETIC BASICS}

The physics behind photophoresis has two regimes.
At low ambient pressure, it can be described by the interaction of particle surfaces with individual gas molecules as mentioned previously (free molecular flow regime).
At high pressure, photophoresis is balancing the thermal creep of gas in a thin layer along the surface of the particle.
While the strength depends on the gas pressure, the effect occurs for any particle which is not in thermal equilibrium with its surroundings.
This is the case for any illuminated particle with warmer and cooler sides.

\subsubsection{Photophoresis at Low Pressure}
In a minimum mass solar nebula, the mean free path of gas molecules at 1 AU is 1 cm \citep{Hayashi1985}. Chondrules are much smaller than the mean free path, especially further out in the asteroid belt region. Therefore, we only consider the case of free molecular flow here. The calculation of the photophoretic force then is reduced to calculating the momentum balance of gas molecules impinging and leaving the particle surface. There are two types of collisions of gas molecules with a surface: diffuse and specular reflections. Specular reflections on average do not alter the particle's momentum. Diffuse reflection refers to the case where the gas molecule is adsorbed onto the surface for a short time and is then ejected in an arbitrary direction. This can change the momentum of a particle since the ejected molecule carries a momentum related to the local surface temperature. The fraction of diffusely reflected molecules, denoted here as $\alpha$, is commonly called the thermal accommodation coefficient. The fraction $\alpha$ is a property of the specific gas species and particle surface and is often considered to be 1.

The surface temperature of an illuminated and absorbing particle is not homogeneous along the surface. Gas molecules leave faster from hot surface parts than from colder parts. The net momentum transfer is given as a surface-integral over the particle of the local momentum induced by interactions at the local temperature \citep{Rohatschek1985, Rohatschek1995, Hidy1970}:
\begin{equation}
    \mathbf{F}=-\frac{1}{2} \oint\limits_{S_{\mathrm{par}}} p \left( 1+\sqrt{1+\alpha\left(\frac{T}{T_\mathrm{gas}}-1\right)} \right) \mathbf{d}\boldsymbol{\sigma} \; . \label{eq:force}
\end{equation}
Here, $T$ is the particle's local surface temperature, and $T_\mathrm{gas}$ and $p$ are the gas temperature and pressure far away from the particle, respectively.
$S_\mathrm{par}$ is the particle's surface.
Our photophoretic force calculations are based on Equation (\ref{eq:force}) where we calculate the temperature distribution at the particle surface for a given light flux $I$ as detailed below.
We further assume that the accommodation coefficient is constant over the particle surface.
Photophoresis is then caused by different surface temperatures which is sometimes called $\Delta T$ photophoresis \citep{Cheremsin2005}.
The case of accommodation coefficients which vary over a particle's surface is discussed in the literature as $\Delta \alpha$ photophoresis \citep{Cheremsin2005, Rohatschek1956A, Rohatschek1956B}.
In extreme cases, i.e., a dichotomy over a particle with two different values of $\alpha$, this can cause a significant photophoretic force. This is also the case even for highly conducting particles if the particle temperature is different from the gas temperature. However, we assume here that the large number of different mineral grains at a chondrule's surface provide enough averaging that $\Delta \alpha$ photophoresis is unimportant and we leave detailed studies on $\Delta \alpha$ effects for the future.

Equation (\ref{eq:force}) is of little practical use for general applications. For this reason, simplifications have been worked out for specific particle types in the past.

\subsection{Photophoretic Forces of Spherical Particles}

Approximations for photophoresis on well-conducting, spherical particles are given in the literature \citep{Rohatschek1985, Beresnev1993}.
These  are  sufficient  for  general  estimates on particle transport. However, \citet{Loesche2012} found that the deviations from the real photophoretic force can be up to a factor of three (depending on the setting) which is not sufficient for accurate estimates, e.g. on the sorting capability.
Therefore, we developed a more accurate and complex but still analytic equation to describe photophoretic forces for homogeneous spherical particles at low pressure. It is \citep{Loesche2012}
\begin{equation}
\begin{split}
 F &= F\left(r,k,\alpha,I,T_\mathrm{gas}^\mathrm{kin}, T_\mathrm{gas}^\mathrm{opt} \right)\\
   &= \left(0.7231-0.1741 e^{-2.180 \frac{r}{k}\mathrm{W/(m^2\,K)}} + 0.4316 e^{-0.9251 \alpha} \right) \text{\boldmath$\cdot$}\\
   & \quad \text{\boldmath$\cdot$} \, \alpha \, \frac{\pi}{6} \, \frac{p}{T_\mathrm{gas}^\mathrm{kin}} \, I \, r^2  \left[ \frac{k}{r}+4\sigma\left(\frac{I}{4\sigma} + \left( T_\mathrm{gas}^\mathrm{opt} \right)^4\right)^{\frac{3}{4}} \right]^{-1}
 \; , \label{eq:forceNumerical}
\end{split}
\end{equation}
where $k$ is the thermal conductivity of the particle, $\sigma$ is the Stefan-Boltzmann constant, $r$ is the particle radius, $I$ is light flux density, and $T_\mathrm{gas}^\mathrm{opt}$ and $T_\mathrm{gas}^\mathrm{kin}$ are the temperatures of the radiation field and the thermal energy of the gas molecules, respectively.
The latter two can be different in optical thin environments.

This equation can also be used for onion shell like particles if an effective thermal conductivity is attributed to the particle \citep{Loesche2012}. We will name this $\varkappa$ here to distinguish it from
$k$-values of the constituents.

\subsection{Non homogeneous, Non spherical Particles}\label{sec:5b}

For homogeneous spheres or onion-shell particles, the photophoretic force is directed along the line of illumination due to symmetry, and the most important particle property to consider is the thermal conductivity $\varkappa$.
For non homogeneous and non spherical particles, the force does not necessarily move a particle in the direction of the light, and the absolute force can also vary with the orientation of a particle in a given illumination field.
The goal here is to evaluate these variations of the photophoretic force (magnitude and direction) for a chondrule.
The basic means to do so is by calculating the force with a given detailed composition using Equation (\ref{eq:force}).
It has to be noted that it is not clear, \textit{a priori}, that  variations in photophoretic forces for different orientations of a particle are small.
Therefore, even though measurements for average thermal conductivities exist for meteorites \citep{Opeil2010, Opeil2012}, these averages of thermal conductivities cannot be used for photophoretic force calculations of chondrules as quantifying the deviations was one major goal of this work.

It is clear that Equation (\ref{eq:force}) is not useful for the description of the average behavior of particles for further use in photophoretic transport models. Eventually, some effective thermal conductivity might be representative for describing photophoresis on chondrules. If so, and if a characteristic size for the (non spherical) chondrule can be given, Equation (\ref{eq:forceNumerical}) can also be used to calculate the average photophoretic force for non spherical particles.

\begin{figure}[h!]
\begin{center}
\includegraphics[height=0.9\columnwidth]{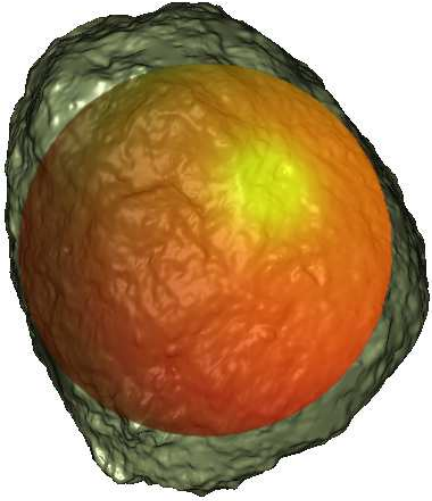}
\caption{Three-dimensional visualization of a chondrule and the largest sphere inscribed centered on the center of shape}
\label{fig:chondruleWithSphereInscribed}
\end{center}
\end{figure}

Variations in photophoretic force can be caused by two aspects, the inhomogeneity and the non sphericity. To estimate either one, we reduce the real chondrules.
To compare the force on a real chondrule to photophoresis on a spherical but inhomogeneous particle, we calculated a spherical approximation of a chondrule by inscribing the largest spherical particle that fits into the chondrule but keep the original inhomogeneity (Figure \ref{fig:chondruleWithSphereInscribed}).

To consider the pure effect of the non sphericity of the chondrule, we also consider particles of chondrule shape that are chemically homogeneous, i.e., attributing them to a single $k$.
This allows us to separate effects of non sphericity and inhomogeneities (if different) on the direction of photophoretic forces and the variations of the absolute force.
In summary, the steps within this paper are to
\begin{itemize}
\item numerically evaluate the photophoretic force on chondrules of a measured shape and composition;
\item numerically evaluate the photophoretic force on spherical particles of chondrule composition;
\item numerically evaluate the photophoretic force on chondrule-shaped but homogeneous particles;
\item quantify deviations in the direction of the photophoretic force with respect to the illumination for the real chondrules, the inhomogeneous sphere approximations and the homogeneous models of chondrule shape;
\item find a suitable average radius of a chondrule with respect to the photophoretic force \\ and
\item find a suitable effective thermal conductivity depending on chondrule composition.
\end{itemize}

\section{FROM THREE-DIMENSIONAL TOMOGRAPHY TO SURFACE TEMPERATURE MAPS}

\subsection{Tomography}
For this study, we used a sample of 19 chondrules from the Bjurb\"ole chondrite (L/LL4-chondrite type).
The heat transfer calculations require the chondrule shape and the composition (thermal conductivity according to mineralogy) of the chondrules.
To obtain accurate digital representations of their surfaces and interior, we used synchrotron X-ray microtomography ($\mu$CT) followed by digital data extraction (e.g., \citet{Friedrich2008, Sasso2009b}).
Chondrules were imaged at a resolution of $5.26\,\mathrm{\mu m/voxel}$ with 30 keV monochromatic X-rays at the 13-BMD synchrotron beamline located at the Advanced Photon Source (APS) of Argonne National Laboratory with the experimental setup presented in \citet{Ebel2007}.

\begin{figure}[h!]
   \centering
   \def\svgwidth{\columnwidth} % sets the image width, this is optional
   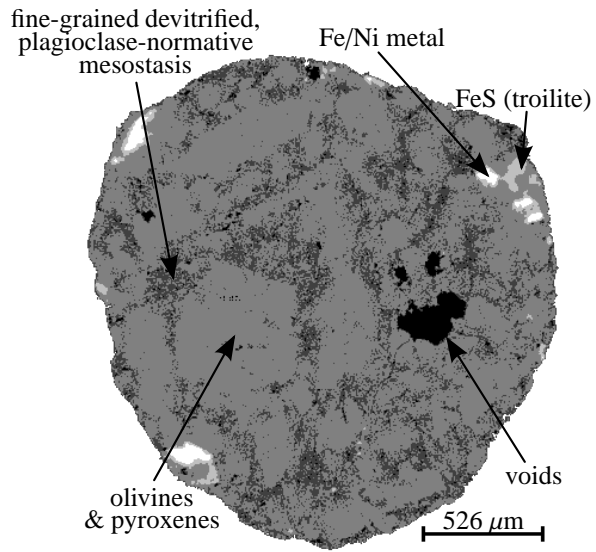
   \caption{Typical X-ray tomogram (contrast enhanced) of a Bjurb\"ole chondrule obtained at a resolution of $5.26\,\mathrm{\mu m/voxel}$. High mean atomic weight materials are bright.}
   \label{fig:ChantalJon}
\end{figure}

To facilitate throughput, we collectively analyzed the chondrules in a specially constructed poly(methyl methacrylate) honeycomb-like receptacle.
Each chondrule was placed within an individual cell for analysis; this facilitated post-examination identification and the retrieval of individual chondrules.
Therefore, all chondrules were analyzed under identical experimental conditions and adequate space was present between them to facilitate later digital separation.
After isolating the volume surrounding each chondrule, we used BLOB3D \citep{Ketcham2005} to obtain abundances of mineralogical components, porosity, bulk volume, and density.
BLOB3D allows for the digital separation, segmentation, and quantification of materials present in digital volumetric datasets such as those produced by our apparatus.
A typical tomographic slice is shown in Figure \ref{fig:ChantalJon}.

\subsection{From Tomography to Three-dimensional Model}\label{sec3}

\begin{figure}[h!]
   \centering
   \includegraphics[width=\columnwidth]{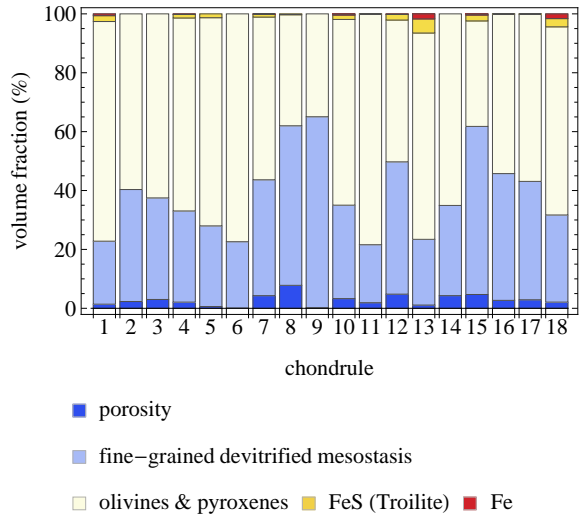}
   \caption{Average modal composition of each chondrule.}
   \label{fig:compounds}
 \end{figure}

After tomographic analysis, a series of slices existed for a given chondrule.
All boundary points separating the  background from the chondrule material were extracted from these slices.
The resulting boundary point cloud was then converted into a geometric object containing parameterized surfaces (called NURBS). For a chondrule's interior, the grayscale values from tomography were mapped to certain minerals. Each chondrule is composed of Fe,Ni-metal, FeS, olivines and pyroxenes, a fine-grained devitrified, plagioclase-normative mesostasis, and voids (Figure \ref{fig:compounds}). Thermal conductivities resembling these materials were used at each point (Table \ref{tab:Material}).
A sample tomogram used for $k$-assignment is shown in Figure \ref{fig:ChantalJon}.
In general, the thermal conductivities of the minerals are functions of the temperature. For $T > 200\,\mathrm{K}$ \citet{Opeil2012} find a $1/T$ dependence for meteoritic material. Furthermore, the thermal conductivities are also dependent on the porosity within the chondrule. In our chondrule set, we do see microcracks in the tomography on the single
voxel level. However, cracks can occasionally span wide nets within chondrules. We assigned a low thermal conductivity of $0.01\,\mathrm{W/(m\cdot K)}$ to any empty space to account for radiative heat transfer within. We also consider the mesostasis to be porous and therefore attribute a low thermal conductivity to the mesostasis independent of the actual mineral conductivity. This is an estimate. According to the temperature dependence of the thermal conductivity the absolute values of the photophoretic force will change with temperature as well. The scaling to account for temperatures at different locations in the solar nebula would be subject to further research and is beyond this study.
Again, possible unresolved microcracks lower the thermal conductivity which is another scaling factor.
However, we consider this approach for attributing a single thermal conductivity as suitable for the evaluation of relative variations of the photophoretic force with orientation due to variations in composition and shape.

\begin{table*}
\footnotesize
\caption{Chondrule Brightness Substitution and Material Properties. Typical Bjurb\"ole olivine and pyroxene have a density of about $3.5$ and $3.4\,\mathrm{\frac{g}{cm^3}}$, respectively. Considering a plagioclase normative mesostasis having about 50\% plagioclase (An10-15) and 50\% mafic silicates a density of about $3.0\,\mathrm{\frac{g}{cm^3}}$ can be estimated. However, in this paper only a stationary problem is solved, so $c_p$ and $\rho$ have not been used yet.}

\begin{threeparttable}
%\begin{tabular}{lccccc} \toprule
%\begin{tabular}{>{\centering\arraybackslash}p{4.3cm} p{2.5cm} p{3cm} p{3cm} p{3cm} p{3cm}} \toprule
\begin{tabular}{L{4.5cm} c C{2.6cm} c >{\itshape}C{2.5cm} >{\itshape}C{2.5cm}} \toprule
\textbf{Material} & \textbf{Gray Shade} & \textbf{Estimated Gray Value Range} $\left( [0,1] \right) $ & \boldmath$k$ in \unboldmath$\mathrm{\frac{W}{m\cdot K}}$ & \boldmath$c_p$ in \unboldmath$\mathrm{\frac{J}{kg\cdot K}}$ & \boldmath$\rho$ in \unboldmath$\mathrm{\frac{g}{cm^3}}$ \\ \midrule
Fe,Ni-metal 						& white			& $>0.7$ 		& 80.4 \tnote{3}		& 447 \tnote{1} 			& 7.96 \tnote{2}\\
FeS (Iron(II)-sulfide, troilite)  	& light gray 	& 0.35-0.7 		& 4 (mean) \tnote{7}	& 588.5 (mean) \tnote{5} 	& 4.61 (normal temp.) \tnote{6}\\
olivine \& pyroxene 				& gray  		& 0.17-0.35 	& 4.6 (mean) \tnote{7}	& 620 (mean) \tnote{4} 		& 3.45\\
fine-grained devitrified,\newline plagioclase-normative mesostasis
									& dark gray 	& 0.1-0.17 		& 0.1 (estimate)		& 530 (estimate) 			& 3.0\\
void 								& black 		& $<0.1$ 		& 0.01 (estimate)		& 1 						& 0.1\\
\bottomrule
\end{tabular}
\begin{tablenotes}
       \item[1] \citet{Halliday2009};
       \item[2] \citet{Tipler2009};
       \item[3] \citet{Tipler2009a};
       \item[4] \citet{Robie1984};
       \item[5] \url{http://webbook.nist.gov/cgi/inchi/InChI\%3D1S/Fe.S}, Link: Condensed phase thermochemistry data
       \item[6] \url{http://webmineral.com/data/Troilite.shtml}
       \item[7] \citet{Clauser1995}.
\end{tablenotes}
\end{threeparttable}
 
\label{tab:Material}
\end{table*}

%\begin{table}
% \begin{tabular}{l >{\itshape}c >{\bfseries}r}
% apple & red & round\\
% melon & green & round\\
% cookie & brown & square\\
% \end{tabular}
% \caption{Some objects, their color and their shape.}
% \end{table}

In the $\mu$CT images, the fine-grained devitrified mesostasis areas appear in darker grayscales than the (in most cases) considerably coarser-grained olivines and pyroxenes (Figure \ref{fig:ChantalJon}).
However, these ``mesostasis areas'' still contain variable abundances of olivine and pyroxenes, also visible in Figure \ref{fig:ChantalJon}, that cannot be appropriately resolved based on instrumental limitations.

\subsection{Calculating Surface Temperature Maps from Three-dimensional Models}\label{sec4}

Chondrule shapes and compositions extracted from tomography volumes were imported into a finite element method software (COMSOL\textsuperscript{\textregistered}).
We then numerically solved heat transfer through the chondrule by solving the stationary heat transfer equation
\begin{equation}
	\boldsymbol{\nabla} \cdot k\boldsymbol{\nabla} T = 0 \; . \label{eq:heatEq}
\end{equation}
As a boundary condition, we included cooling through thermal emission in all directions and heating through absorption of illumination from a given direction.
We assumed an emissivity for thermal radiation of $1$.
For a plane wave light source (light flux $I*\mathbf{e}_I$), these boundary conditions are 
\begin{equation}
\mathbf{n} \cdot k 	\boldsymbol{\nabla}T = I\,\mathbf{e}_I \cdot \mathbf{n}-\sigma(T^4-T_\mathrm{gas}^4) \; , \label{eq:boundaryCond}
\end{equation}
with $\mathbf{n}$ denoting an outwards oriented surface normal.
Corresponding to lab experiments (for comparison in Paper 2), the light flux was set to $I = 20\,\mathrm{kW/m^2}$ and the environment's temperature was $T_\mathrm{gas} = 293\,\mathrm{K}$.
The chondrules heated up until they were in radiative equilibrium with their environment.

\subsubsection{Numericals}
This method was already used and extensively tested in \citet{Loesche2012}. 
COMSOL uses tetrahedral meshes consisting of finite elements (edges) of certain lengths. In the highest resolution that can be calculated, the mean size of a mesh cell is $23\,\mathrm{\mu m}$.  In relation to a tomography resolution of $5.26\,\mathrm{\mu m}$, this is comparable but refinements of the mesh can slightly alter the fine structure simulated.

\begin{figure}[h!]
   \centering
   \includegraphics[width=\columnwidth]{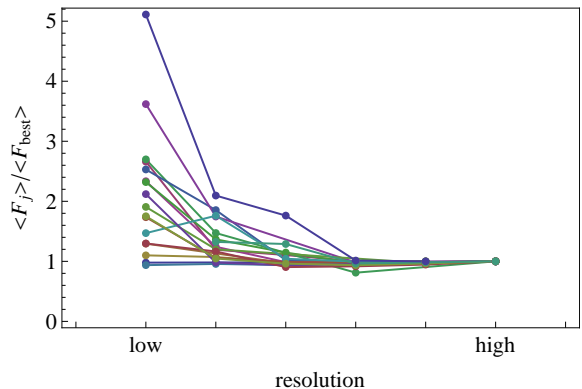}
   \caption{Average photophoretic force for different numerical resolutions and chondrules, normalized to the value achieved with the highest resolution possible. Values converge at the higher mesh resolutions.}
   \label{fig:convergenceForce}
\end{figure}

Numerically, calculations show a very strong convergence in terms of both orientation of the photophoretic force and absolute force value with decreasing mesh size (Figure \ref{fig:convergenceForce}). Variations between the two highest resolution meshes are on the order of 2\% of the average absolute force (discussed below) which we attribute to the slight changes in resolving the tomographic details.

\subsubsection{Parameter Space}

Light flux and ambient temperature were chosen to match experimental values. The light flux corresponds to a radial distance of $0.26\,\mathrm{AU}$ for a 1 solar luminosity star when  opacity is not considered. Since the asteroid belt is at about $3\,\mathrm{AU}$, a light flux of $152\,\mathrm{W/m^2}$ would be expected. The ambient temperature was chosen to match experimental conditions for comparison (room temperature). In protoplanetary disks, temperatures vary between tens of Kelvin to sublimation temperatures.
It might be worth noting that the typical difference in temperature over a $1\,\mathrm{mm}$ diameter particle with thermal conductivity of $0.5\,\mathrm{W/(m\cdot K)}$ at the given light flux is $37\,\mathrm{K}$ between the warm and cold side. However, while we use specific values here, we give only relative outcomes, and the absolute values are not important in the context of this paper. To attribute a specific photophoretic force for a given setting of a solar nebula, Equation (\ref{eq:forceNumerical}) can be used if particle size and effective thermal conductivity are fixed as detailed below.

\section{PHOTOPHORETIC FORCES}\label{sec:Forces}

\subsection{Full Chondrules}\label{sec:Forces:Chondrules}

The resulting photophoretic forces were calculated from a surface's temperature field by Equation (\ref{eq:force}).
In the free molecular flow regime, the absolute force depends linearly on the absolute ambient pressure.
Where absolute numbers are necessary, below we considered the ratio between photophoretic force over pressure.

\begin{figure}[h!]
   \centering
   \includegraphics[width=\columnwidth]{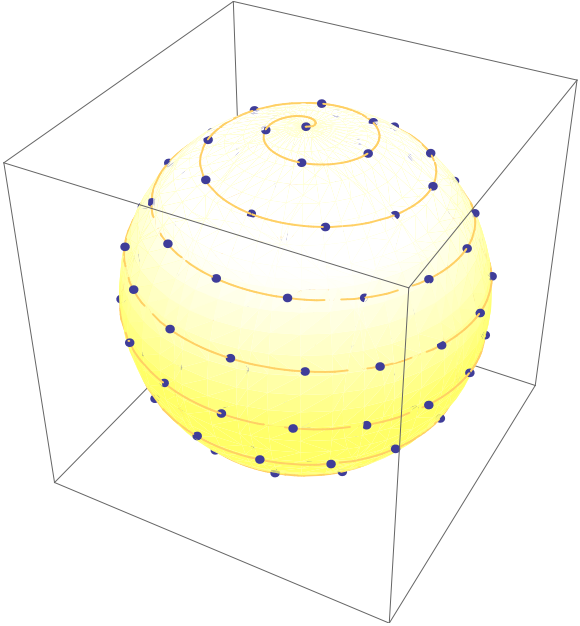}
   \caption{Visualization of the 100 different orientations of incident light with respect to the chondrule. Points are distributed evenly over the sphere to a good approximation by placing them on a spiral.}
   \label{fig:spiral}
\end{figure}

We calculate the photophoretic force for real chondrules with their detailed shape and composition for $N=100$ different orientations with respect to the light source.
The different orientations of the light source are distributed evenly surrounding the chondrule by placing them along a spiral \citep{Rakhmanov1994}
\begin{eqnarray}
 \phi\left(N,t\right)&=&2 \lfloor N^{0.485} \rfloor \arccos \left(\frac{1+N-2 t}{1-N}\right) \;, \label{eq:spiral2b}\\
 \theta\left(N,t\right) &=&\arccos \left(\frac{1+N-2 t}{1-N}\right) \;, \label{eq:spiral2c}
\end{eqnarray}
where $\phi$ and $\theta$ denote the spherical coordinates of incident light and the integer parameter $t$ addresses each point.
The spiral is seen in Figure \ref{fig:spiral}.

\subsubsection{Angular Deviation to Incident Light Direction}

\begin{figure}[ht!]
   \centering
   \includegraphics[width=\columnwidth]{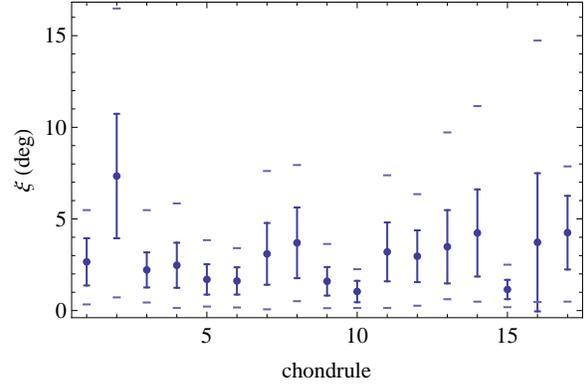}
   \caption{Full chondrules: angular deviation of the photophoretic force from the direction of incident light for 100 different orientations. Error bars are standard deviations, line markers indicate maximum and minimum values.}
   \label{fig:angles1}
\end{figure}

The angular deviation, $\xi$, of the photophoretic force from the direction of incident light is shown in Figure \ref{fig:angles1}.
Here, error bars represent standard deviations ($1\sigma$) over the 100 orientations and line markers indicate maximum and minimum values.

The total average over all chondrules is $\xi = 3.0\,^{\circ} \pm 1.5\,^{\circ}$ (standard deviation).
The force deviates only modestly with illumination direction. Nevertheless, it implies a small sideward motion of illuminated chondrules.
Such a sideward motion is indeed visible in drop tower experiments. Details are discussed in Paper 2.

\subsubsection{Variation of Absolute Forces}

\begin{figure}[ht!]
   \centering
   \includegraphics[width=\columnwidth]{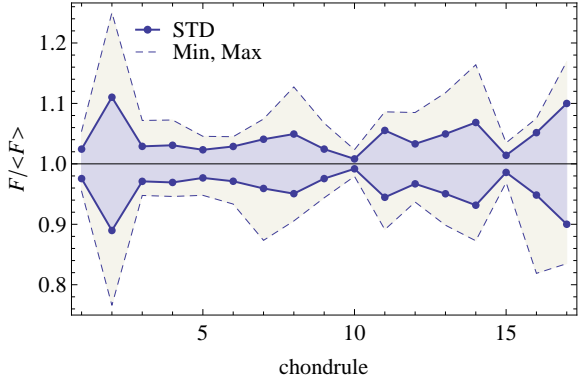}
   \caption{Deviations of the photophoretic force with random orientation from the average force on each chondrule; standard deviations and minimum and maximum deviations.}
   \label{fig:absolute1}
\end{figure}

With inhomogeneity and non-sphericity the absolute force varies with orientation of the light source. The relative deviations of the absolute force to the average photophoretic force are shown in Figure \ref{fig:absolute1}. Typical deviations are a few percent. The average is $\pm 4.17$\%.

\subsection{Spherical Chondrules}\label{sec:Forces:Spheres}

Variations of photophoretic force in  direction and strength might be attributed to inhomogeneities of mineral distribution throughout the particle or the shape of the chondrules or a combination of these factors. This can be studied by comparing the results to calculations with spherical particles. To estimate the effect of chondrule composition, we reduce real chondrules to spherical models by inserting spheres of maximum size (centered at the center of shape) which still completely fit inside. Otherwise, we hold the  composition within the volume enclosed by the sphere as in the original chondrule and calculate the photophoretic force as described above. In some cases, voids at the surface of these inscribed spheres existed. To simulate a spherical particle, we assigned olivine and pyroxenes to the voids

\subsubsection{Angular Deviation to Incident Light Direction}

\begin{figure}[h!]
   \centering
   \includegraphics[width=\columnwidth]{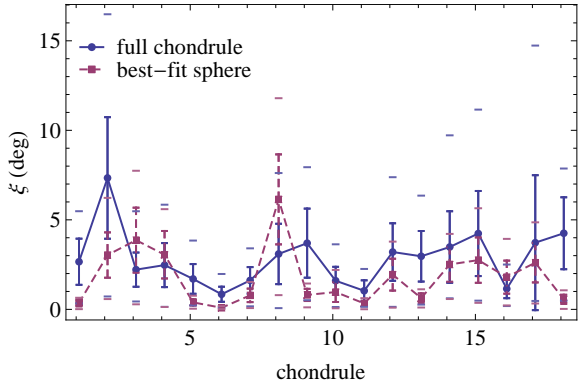}
   \caption{Comparison of variation in the photophoretic force direction between spherical and real chondrules.}
   \label{fig:comparisonAnglesSphereChondrule}
\end{figure}

The total average of all angles for inhomogeneous but spherical particles is $1.81\,^{\circ} \pm 1.59\,^{\circ}$. Fig. \ref{fig:comparisonAnglesSphereChondrule} shows a similar impact of inhomogeneity and asphericity on the scattering angle, $\xi$.
%A comparison between sphere and real chondrule is shown in Fig. \ref{fig:comparisonAnglesSphereChondrule}.

\subsubsection{Variation of Absolute Forces}

\begin{figure}[h!]
   \centering
   \includegraphics[width=\columnwidth]{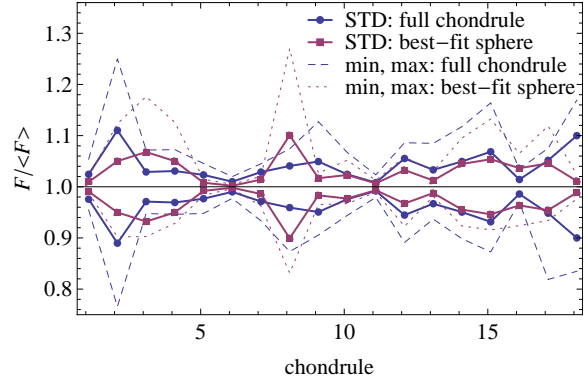}
   \caption{Comparison of all normalized force deviations between spherical and real chondrules.}
   \label{fig:comparisonForcesSphereChondrule}
\end{figure}

The average variation of the force to the average force ratio is $3.25$\%.
A comparison to the real chondrules is given in Figure \ref{fig:comparisonForcesSphereChondrule}.

The spherical particles with chondrule composition show deviations of the photophoretic force in direction and absolute value on the same order as the deviations for the real chondrule calculations.
Therefore, the non-homogeneous composition is taken as a significant parameter influencing the photophoretic force.

\subsection{Homogeneous Particle of Chondrule Shape}\label{sec:Forces:Geometries}

To get the full picture the dependence of shape was studied as well.
We considered particles of chondrule shape but attributed certain constant thermal conductivities to their interior.
Again, the deviation of the photophoretic force from the direction of incident light as well as the variation in absolute value was calculated .
We only briefly discuss one chondrule-like particle here.

\subsubsection{Angular Deviation to Incident Light Direction}

\begin{figure}[h!]
   \centering
   \includegraphics[width=\columnwidth]{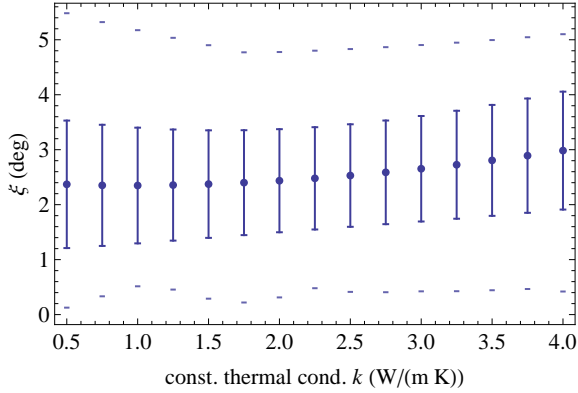}
   \caption{Angular deviation between photophoretic force and direction of illumination for a particle of chondrule shape at different thermal conductivities $k$.}
   \label{fig:angles3Stat}
\end{figure}

Figure \ref{fig:angles3Stat} shows the average angle $\xi$ between $\mathbf{e}_I$ and $\mathbf{F}$ for the chondrule %\textit{18} 
with different thermal conductivities. The average of all angles is $2.54\,^{\circ} \pm 1.00\,^{\circ}$.

\subsubsection{Variation of Absolute Forces}

\begin{figure}[h!]
   \centering
   \def\svgwidth{\columnwidth} % sets the image width, this is optional
   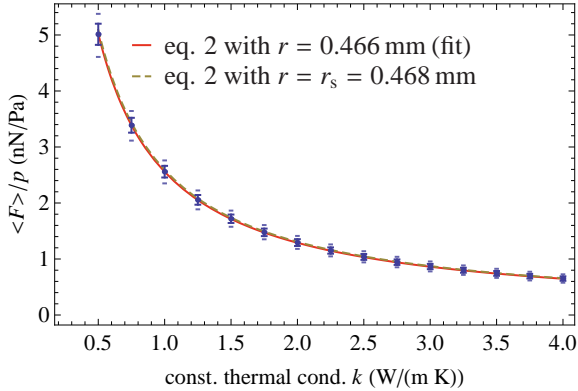
   \caption{Photophoretic force over pressure for a chondrule-shaped particle at homogeneous thermal conductivities $k$. Overplotted are functions according to Equation (\ref{eq:forceNumerical}) with fitted radius $r$ and equivalent volume sphere radius, only deviating $0.45$\% from the fit (both function graphs are almost congruent).}
   \label{fig:force3Stat}
\end{figure}

Average photophoretic forces ($F/p$) and their deviations are shown in Figure \ref{fig:force3Stat}. % and \ref{fig:force3Stat2}.
For spherical particles, Equation (\ref{eq:forceNumerical}) can be used to calculate photophoretic forces. Here, we can use this equation to test if a given size can be used to describe a chondrule as sphere.
We find that the radius of a volume equivalent sphere describes a chondrule-shaped particle very well.
Such a function (Equation (\ref{eq:forceNumerical})) has been fitted for a radius and is overplotted in Figure \ref{fig:force3Stat} for one example.

%\begin{figure}[h!]
%   \centering
%   \def\svgwidth{\columnwidth} % sets the image width, this is optional
%   \input{forceStatisticsHomogeneousChondrule_Gerhard.eps_tex}
%   %\includegraphics[width=\columnwidth]{forceStatisticsHomogeneousChondrule_Willi.eps}
%   \caption{Photophoretic force over pressure for chondrule \textit{Ge} shaped particle at    homogeneous thermal conductivities $k$.}
%   \label{fig:force3Stat2}
%\end{figure}

\begin{figure}[h!]
   \centering
   \includegraphics[width=\columnwidth]{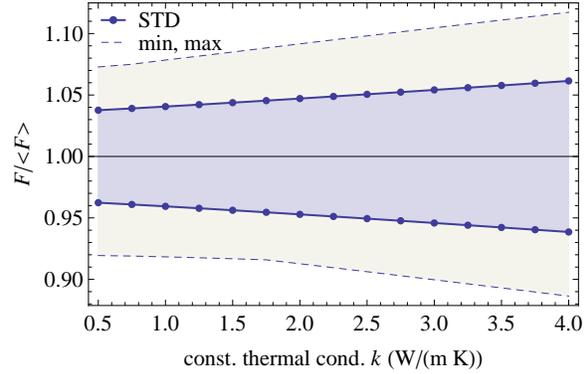}
   \caption{Relative deviations of the photophoretic forces for random orientations of a chondrule-shaped particle for given thermal conductivities $k$.}
   \label{fig:force3StatNormalized}
\end{figure}

As with the previous calculations for real chondrules and spherical particles with chondrule composition, the photophoretic force for homogeneous chondrule shape particles has a certain variation of a few \% slightly increasing with increasing $k$ (see Figure \ref{fig:force3StatNormalized}).

\begin{figure}[h!]
   \centering
   \def\svgwidth{\columnwidth} % sets the image width, this is optional
   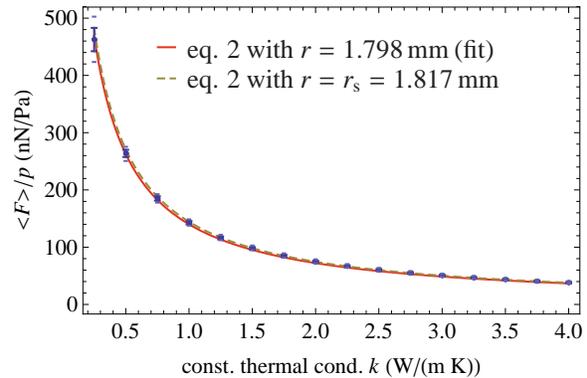
   \caption{Photophoretic force over pressure for an ellipsoid with half-axes $(1,2,3)\,\mathrm{mm}$ for different homogeneous thermal conductivities $k$. Both function graphs are very close to each other.}
   \label{fig:forceEllipsoidStat}
\end{figure}

To test the hypothesis that the equivalent volume sphere radius $r_{\mathrm{s}}$ enters in the calculations, we considered an ellipsoid as another particle shape. The mean force for all different thermal conductivities, even in this case of a very aspherical particle of an ellipsoid with half-axes $(1,2,3)\,\mathrm{mm}$ could also be well approximated by a sphere with equivalent volume radius $r_{\mathrm{s}}=1.82\,\mathrm{mm}$ (see Figure \ref{fig:forceEllipsoidStat}).

\section{PHOTOPHORETIC PROPERTIES OF CHONDRULES}\label{sec:Properties}

The calculations using real chondrule parameters, spherical particles of chondrule composition, and homogeneous particles of chondrule shape show that variations in direction and strength within a few percent occur.
Deviations from perfect spheres, the shape as well as the non-uniform composition, induce variations of the same order.

Overall, the variations are rather small, and for general transport models, it might be sufficient to use average values.
We determine these averages from the composition and as calculated analytically from Equation \ref{eq:forceNumerical} as follows.

\begin{figure}[h!]
   \centering
   \def\svgwidth{\columnwidth} % sets the image width, this is optional
   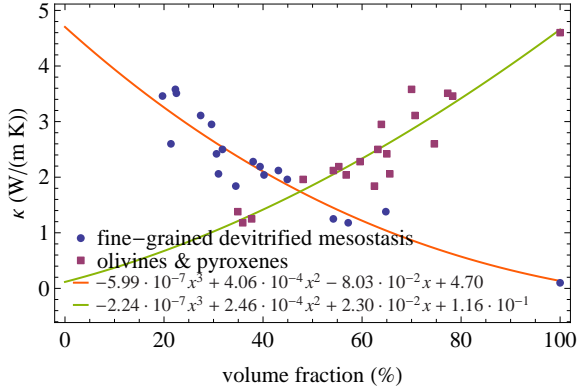
   \caption{Effective thermal conductivity $\varkappa$ over the percentage of fine-grained devitrified mesostasis and olivines and pyroxenes in a chondrule; lines are polynomial fits of third order.}
   \label{fig:kappaChondruleOverCompoundsFit}
\end{figure}

Homogeneous chondrule-like or ellipsoidal particles (see Section \ref{sec:Forces:Geometries}) could be well described by a sphere with the radius of an equivalent volume sphere. For actual chondrules, the effective thermal conductivity $\varkappa$ is the value at which the mean photophoretic force $\langle F \rangle$ of the calculated sphere fits the calculated force of the real chondrule (Equation (\ref{eq:forceNumerical})).
Figure \ref{fig:kappaChondruleOverCompoundsFit} shows the effective thermal conductivity over volume fraction of mesostasis and olivine and pyroxenes.
These are the two main components and differ largely in their individual thermal conductivities.
There is a strong correlation.

\begin{figure}[h!]
   \centering
   \def\svgwidth{\columnwidth} % sets the image width, this is optional
   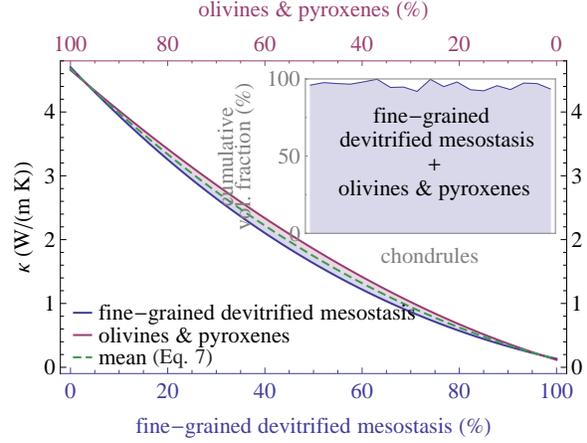
   \caption{Fits from Figure \ref{fig:kappaChondruleOverCompoundsFit}, but olivine and pyroxene fraction now inverted from left to right. Inset shows total mass fraction of these two phases.}
   \label{fig:kappaChondruleOverTwoPhases}
\end{figure}

Figure \ref{fig:kappaChondruleOverTwoPhases} has the olivine and pyroxene fit reverted. Both can be described by one average third order polynomial as
\begin{equation}
	\varkappa_{} = -1.87\cdot 10^{-7} x^3 + 2.94\cdot 10^{-4} x^2 - 7.22\cdot 10^{-2} x + 4.54 \; , \label{eq:kappaChondrule}
\end{equation}

where $x$ is the mesostatis-share in the corresponding two-phase system. Taken together, we find that in spite of detailed compositional differences and shapes we can well describe the chondrule by an effective thermal conductivity only based on the ratio between the two primary phases (Equation (\ref{eq:kappaChondrule})).
Hence, other chondrules of the same type (L/LL4-chondrite type) only need to undergo a determination of the mesostasis and olivines and pyroxenes share as well as the volume, so that $\varkappa$ and eventually $\langle F \rangle$ can be calculated.

\begin{table*}
\scriptsize
\centering
\caption{chondrule properties}

\begin{threeparttable}
\begin{tabular}{>{\bfseries}C{.4cm} C{.7cm} C{.6cm} C{.8cm} C{.6cm} C{.6cm} C{.5cm} C{.5cm} >{\bfseries}C{.5cm} C{.4cm} >{\bfseries}C{.5cm} C{1.1cm}} \toprule
\multicolumn{2}{c}{} & \multicolumn{5}{c}{\textbf{Compounds} (\%)} & \multicolumn{3}{c}{\textbf{Radii} $(\mathrm{mm})$} & \multicolumn{2}{c}{\textbf{Eff. Thermal}} \\
%\cmidrule (l){11-12}
\multicolumn{7}{c}{} & \multicolumn{3}{c}{\tiny (Ref. Point: Center of Shape)} & \multicolumn{2}{c}{\textbf{Conductivity} $\boldsymbol{\varkappa} \left( \mathrm{\frac{W}{m\cdot K}}\right)$} \\
\cmidrule (l){3-7} \cmidrule (l){8-10} \cmidrule (l){11-12}
 Sample & Total\newline Volume ($\mathrm{mm^3}$) & Porosity & Fine-\newline Grained\newline Devitrified\newline Mesostasis & Olivines and \newline Pyroxenes & FeS (Troilite) & Fe,Ni-metal & $r_{\max}$ & $\boldmath r_{\text{s}}$ & $r_{\min}$ & Mean & STD Interval \\ %\tnote{c}
\midrule
 \text{1} & 5.065 & 1.4 & 21.4 & 74.6 & 2.0 & 0.6 & 1.36 & 1.065 & 0.82 & 2.60 & 2.54 - 2.67 \\
 \text{2} & 0.020 & 2.3 & 38.0 & 59.6 & 0.0 & 0.0 & 0.24 & 0.167 & 0.09 & 2.28 & 2.06 - 2.57 \\
 \text{3} & 0.096 & 3.0 & 34.5 & 62.5 & 0.0 & 0.0 & 0.36 & 0.284 & 0.20 & 1.84 & 1.79 - 1.90 \\
 \text{4} & 0.398 & 2.1 & 31.0 & 65.6 & 1.2 & 0.2 & 0.56 & 0.456 & 0.34 & 2.06 & 2.00 - 2.13 \\
 \text{5} & 0.369 & 0.6 & 27.4 & 70.7 & 1.3 & 0.0 & 0.53 & 0.445 & 0.38 & 3.11 & 3.04 - 3.19 \\
 \text{6} & 0.227 & 0.1 & 22.5 & 77.3 & 0.0 & 0.0 & 0.43 & 0.379 & 0.28 & 3.51 & 3.47 - 3.54 \\
 \text{7} & 0.189 & 4.3 & 39.4 & 55.3 & 1.0 & 0.1 & 0.41 & 0.356 & 0.25 & 2.19 & 2.13 - 2.26 \\
 \text{8} & 0.122 & 7.8 & 54.2 & 37.7 & 0.3 & 0.0 & 0.39 & 0.308 & 0.17 & 1.25 & 1.20 - 1.30 \\
 \text{9} & 0.186 & 0.2 & 64.8 & 34.9 & 0.0 & 0.0 & 0.51 & 0.354 & 0.24 & 1.38 & 1.32 - 1.46 \\
 \text{10} & 0.226 & 3.3 & 31.8 & 63.2 & 1.5 & 0.4 & 0.53 & 0.378 & 0.27 & 2.50 & 2.44 - 2.56 \\
 \text{11} & 0.186 & 1.9 & 19.7 & 78.3 & 0.1 & 0.0 & 0.41 & 0.354 & 0.27 & 3.46 & 3.44 - 3.49 \\
 \text{12} & 0.766 & 4.8 & 44.9 & 48.1 & 1.9 & 0.2 & 0.73 & 0.568 & 0.32 & 1.96 & 1.86 - 2.08 \\
 \text{13} & 0.535 & 1.1 & 22.3 & 70.0 & 4.8 & 1.7 & 0.62 & 0.504 & 0.39 & 3.58 & 3.46 - 3.70 \\
 \text{14} & 0.063 & 4.3 & 30.6 & 65.0 & 0.0 & 0.0 & 0.33 & 0.246 & 0.18 & 2.42 & 2.30 - 2.54 \\
 \text{15} & 0.036 & 4.7 & 57.2 & 35.9 & 2.0 & 0.4 & 0.29 & 0.205 & 0.14 & 1.18 & 1.11 - 1.27 \\
 \text{16} & 0.068 & 2.7 & 43.1 & 54.2 & 0.1 & 0.0 & 0.29 & 0.253 & 0.17 & 2.12 & 2.09 - 2.15 \\
 \text{17} & 0.182 & 2.9 & 40.2 & 56.8 & 0.1 & 0.0 & 0.45 & 0.351 & 0.22 & 2.04 & 1.94 - 2.15 \\
 \text{18} & 0.429 & 2.1 & 29.6 & 63.9 & 2.8 & 1.6 & 0.66 & 0.468 & 0.33 & 2.95 & 2.68 - 3.28 \\
\bottomrule
\end{tabular}
%\begin{tablenotes}
%       \item[a] mean of compound values
%       \item[b] sphere with $r=r_{\mathrm{min}}$
%       \item[c] radius of sphere with the same volume
%       \item[d] calculated by solving Eq. \ref{eq:forceNumerical} for $\kappa$ for each force value
%       \item[e] calculated by solving Eq. \ref{eq:forceNumerical} with mean force for $\kappa$
%\end{tablenotes}
\end{threeparttable}

\label{tab:results}
\end{table*}

%\begin{table}
% \begin{tabular}{l >{\itshape}c >{\bfseries}r}
% apple & red & round\\
% melon & green & round\\
% cookie & brown & square\\
% \end{tabular}
% \caption{Some objects, their color and their shape.}
% \end{table}

The results of the calculations are summarized in Table \ref{tab:results}.

\begin{figure}[h!]
   \centering
   \def\svgwidth{\columnwidth} % sets the image width, this is optional
   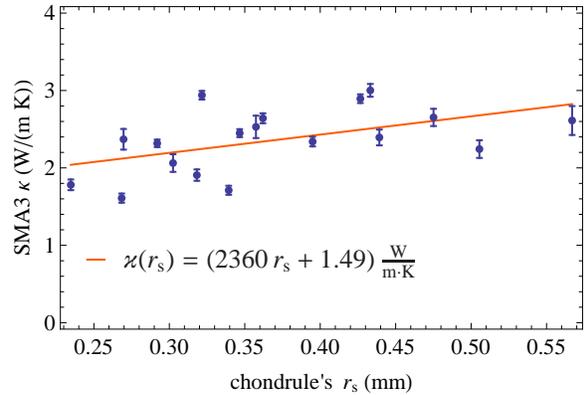
   \caption{Cyclic SMA3-correlation (simple moving average of third order) of effective thermal conductivity $\varkappa$ for real chondrules and their volume equivalent radius $r_\mathrm{s}$.}
 \label{fig:kappas4OverRmean}
\end{figure}

While this can now be applied more easily in general transport models, one immediate application to see here is the correlation between the two parameters quantifying the photophoretic force -- thermal conductivity and size. This is shown in Figure \ref{fig:kappas4OverRmean}.

There is a slight trend that the thermal conductivity varies with size which allows a size sorting according to photophoresis at low pressures \citep{Loesche2012, Wurm2006}, but this depends on the disk model and is subject to further research.

\section{Conclusions}

We currently consider photophoresis to be one of the major forces acting on particles in optically thin regions of protoplanetary disks. This is especially true at the inner disk edge where photophoresis is responsible for local transport \citep{Haack2007, Loesche2012, Wurm2009}.
As detailed in the preceding sections we quantified the photophoretic force on chondrules in this work. We found that the photophoretic force on chondrules can be well approximated on average by Equation (\ref{eq:forceNumerical}) for spherical particles. A radius of a sphere with equivalent volume must be used. One effective value can be used for thermal conductivity. For the chondrule sample studied, this can be calculated from the fraction of low porosity material (mesostasis) as quantified in Equation (\ref{eq:kappaChondrule}). 

It has to be noted that the specific value of $\varkappa$ depends on the choice of the thermal conductivities attributed to the different phases, i.e., fine-grained devitrified mesostasis having $k=0.1$ W/(K m), which is not known to higher accuracy and which might vary from chondrule to chondrule and with temperature. Additionally, microcracks which are smaller than the resolution of the tomography, could not be taken into account. Such microcracks will severely reduce the thermal conductivity, which will be beneficial for photophoresis.
However, this will not change the basic findings of this study and on average will only shift the effective thermal conductivity systematically.

In total, the results allow an analytical treatment of photophoretic forces on chondrules in different applications if their volumes and compositions are known. The calculations of the photophoretic force are based on a parameter set chosen to match with the experimental data in microgravity experiments. This means that parameters like irradiation, gas pressure, or gas temperature do not match with values expected for protoplanetary disks. However, these can easily be rescaled by Equation (\ref{eq:forceNumerical}). For example a chondrule like sample no. 10 is exposed to a photophoretic force twice that of gravity, if it is located in direct sun light at 0.1 AU in a Minimum Mass Solar Nebula \citep{Hayashi1985}.

Independent of the specific surface morphology or mineral distribution within the chondrule the mean absolute forces vary on average by $4.17$\%, depending on orientation. Inhomogeneity and shape both have a similar influence.

The direction of the photophoretic force for chondrules is centered  on the direction of the incident light with average deviations of $3.0^\circ \pm 1.5^\circ$.
Such deviations will induce small sideways forces and motions.
We did not yet quantify torques and rotation induced by photophoresis as seen in experiments by \citet{Eymeren2012}.

In Paper 2, we report experimental measurements of the photophoretic
force on the same set of chondrules and compare the results to the calculations given
in this paper.

\section{Acknowledgements}
Part of this work is funded by the Deutsche Forschungsgemeinschaft within the priority program SPP 1385. Access to the related drop tower experiments has been granted by ESA.

Special thanks to the FEMET GmbH for placing Geomagic Studio 12\textsuperscript{\textregistered} at our disposal.

J.M.F. would like to thank the Fund for Astrophysical Research for assistance in the acquisition of computer equipment used for portions of this study and NASA's Origin of Solar Systems (OSS) Program grant NNX10AH336 (Co-I J.M.F.) for additional support. Portions of this work were performed at GeoSoilEnviroCARS (Sector 13), Advanced Photon Source (APS), Argonne National Laboratory. GeoSoilEnviroCARS is supported by the National Science Foundation -- Earth Sciences (EAR-1128799) and Department of Energy -- Geosciences (DE-FG02-94ER14466). Use of the Advanced Photon Source was supported by the U. S. Department of Energy, Office of Science, Office of Basic Energy Sciences, under Contract No. DE-AC02-06CH11357.

We thank the anonymous referee for his review of our manuscript which we think helped a lot to improve it.

%---Literaturverzeichnis--------------------------------------------------------------------------------------------------
\appendix
\bibliographystyle{apalike} % style apalike or apacite
\bibliography{Referenzen} % your references <filename>.bib

\end{document}